\documentstyle[epsfig]{l-aa}

\begin{document}

\newcommand{\logVM}{\log{V_{\rm M}}}


\thesaurus{11(11.19.2; 11.04.1; 12.12.1)}
\title{{\em Research Note \vspace{0.8cm}\\ } 
            The radial space distribution of KLUN-galaxies up to 200~Mpc: 
            incompleteness or evidence for the behaviour predicted by
	    fractal dimension $\approx$ 2?}
\author{P.~Teerikorpi\inst{1}, M.~Hanski\inst{1}, G.~Theureau\inst{2},
Yu.~Baryshev\inst{3}, G.~Paturel\inst{4}, L.~Bottinelli\inst{2,5}, 
L.~Gouguenheim\inst{2,5}}
\offprints{}
\institute{Tuorla Observatory, FIN-21500 Piikki\"{o}, Finland
\and
 Observatoire de Paris-Meudon, CNRS URA1757, F-92195 Meudon
Principal Cedex, France
\and
 Astronomical Institute of the Saint-Petersburg University,
198904 St. Petersburg, Russia
\and
 Observatoire de Lyon, F-69230 Saint-Genis Laval, France
\and
 Universit\'{e} Paris-Sud, F-91405 Orsay, France}
\date{Received \hspace*{3cm}; accepted}
\maketitle
\markboth{Teerikorpi et al.: The radial space distribution of KLUN-galaxies}{ }

\begin{abstract}
We have studied using the KLUN sample of 5171 spiral
galaxies having Tully-Fisher distance moduli, the average radial
space distribution of galaxies out to a distance of about 200~Mpc
(for $H_0=50$~km~s$^{-1}$~Mpc$^{-1}$).  
One motivation came from the debate on the fractal
dimension $\mathcal{D}$ and maximum fractality scale $\lambda_{\rm max}$
of the large-scale galaxy distribution (Davis 1997, Guzzo 1997, 
Pietronero et al.\
1997). A specific recent study is the 3-dimensional correlation
analysis of the all-sky LEDA data base by Di Nella et al.\ (1996)
who concluded that the galaxy distribution is fractal up to
scales of at least 300~Mpc, with fractal dimension $\approx 2.2$.
One would expect to see a signal of this 
result in the radial space
distribution of the all-sky KLUN sample.  We have studied this
question with a new method based on photometric
TF distances, independent of redshift, 
 to construct the number density distribution.

Our main results are: \begin{enumerate}
\item{While scattered below about 20 Mpc, at larger distances the radial
distribution starts to follow, in terms of distance modulus $\mu_{\rm TF}$,
the law $\log{N} = (0.46 \pm 0.01) \mu + $const., using diameter TF relation,
and $\log{N} = (0.40 \pm 0.01) \mu + $const. for magnitudes.
These are the predictions based on fractal dimensions 2.3 and 2.0, 
respectively. These radial density gradients are valid up to the limits 
of KLUN, or about 200 Mpc.}
\item{We have tried to understand the derived radial density behaviour as a 
result of some bias in KLUN or our analysis, however, without success.
Numerical simulations have shown that the method itself works,
though it somewhat underestimates the radial distribution exponent.
If the density law is caused by incompleteness in the diameter limited
KLUN sample, then the incompleteness should start at widely different
angular diameters $d_{25}$ for different values of rotation parameter $\logVM$,
which would be quite unexpected. On the other hand, if the derived
distribution is correct, the completeness is good down to $d_{25} = 1'.6$,
as originally intended and previously concluded.}
\item{If correlation studies favoring long scale 
fractality (200 Mpc or more) and $\mathcal{D} \approx$ 2
are correct, the position of our Galaxy
would be close to average in the Universe, with the galaxy density decreasing
around us according to the expected law (Mandelbrot 1982).}
\end{enumerate}


\end{abstract}

\section{Introduction}

The present work 
is motivated by the longstanding problem in observational cosmology,
concerning the structure of galaxy distribution: is the Universe
homogeneous or fractal on large scales, and what is the scale
where homogeneity begins? Paturel et al.\ (1994) and
Witasse \& Paturel (1997) left open the possibility that the 
apparent partial incompleteness in some galaxy samples could 
actually reflect inhomogeneity or fractal distribution.

The completeness of the KLUN-sample of spiral galaxies was
studied by Paturel et al.\ (1994) from the distribution of the
angular sizes ($\log{d_{25}}$, with isophotal 25 mag/arcsec$^2$ 
diameters $d_{25}$ expressed in units of
$0'.1$).  Separating the sample into parts with  and without the plane of
the Local Supercluster, it was concluded (on the assumption of
homogeneity outside of the LS) that the sample is almost complete 
down to $\log{d_{25}} = 1.2$ (or $d_{25} = 1'.6$).  
In fact, it is a general problem, how to determine the completeness of
a sample, if there is no a priori quarantee that the galaxy counts in a 
complete sample should follow, say, the distribution predicted by a homogeneous
distribution of galaxies.

We continue here the study of
incompleteness (the knowledge of which is crucial for our
applications of KLUN), and attempt simultaneously to determine the
all-sky average radial distribution of galaxies.  This is done by
a method which utilizes the TF distance moduli available for
every KLUN galaxy.
As the question of the radial distribution is closely connected with the 
problem of the scale and dimension of fractal distribution,  we give
below a short resum\'{e} of the state of this field.

The early history of determination of the fractal dimension and
the maximum scale of fractality is described by Peebles (1980):
the main result was that $\mathcal{D} \approx$ 1.2 and $\lambda_{max} 
\leq 20$ Mpc\footnote{All distances in this paper are based on
$H_0=50$~km~s$^{-1}$~Mpc$^{-1}$.}.
The first evidence for a universal fractal distribution of galaxies with
fractal dimension $\mathcal{D} \approx$ 2 was given by Baryshev (1981)
from number counts and virial mass distribution arguments. The same
$\mathcal{D} \approx$ 2 inside the Virgo supercluster 
was indicated by Klypin et al.\
(1989) from analysis of fractal properties of superclusters, with
$\lambda_{max} < 40$ Mpc. Later in several works a double power law
behaviour was found so that at distances $r \leq 6$ Mpc $\mathcal{D} \approx$ 1.2,
but for 6 Mpc $< r < 60$ Mpc there is $\mathcal{D} \approx$ 2.2 (Einasto 1991,
Guzzo et al.\ 1991, Calzetti et al.\ 1992, Einasto 1992, Guzzo 1997). 

In recent years several extensive galaxy redshift surveys have become
available (CfA1, CfA2, SSRS1, SSRS2, Perseus-Pisces, LEDA, LCRS, IRAS, etc.,
see the review by Sylos Labini et al.\ 1998a). Detailed analyses of the surveys
have been made by different quantitative statistical methods which have
revealed the fractal structure of the 3-dimensional galaxy distribution
at the scales corresponding to the surveys. The fractality implies that
around any galaxy the number density behaves with radius $r$ as 
$n(r)= k r^{({\mathcal{D}}-3)}$ for $r\leq \lambda_{max}$.

However, considerable debate has arisen on the value of fractal dimension
$\mathcal{D}$ and the maximum correlation length $\lambda_{max}$
(see e.g.\ Guzzo 1997, Sylos Labini et al.\ 1998b). One may even speak
about two schools. According to the first one (see Davis 1997), using the
standard correlation function analysis, fractal structure exists within
the scale interval 0.02 Mpc $< r <$ 20 Mpc with dimension 
$\mathcal{D} \approx$ 1.2, while on the scales larger than 40 Mpc the Universe
becomes homogeneous. According to the second school (see Pietronero
et al.\ 1997, Sylos Labini et al.\ 1998a), using a more general
statistical approach which includes the correlation function method
as a particular case, the fractal structure exists at least up to 200 Mpc
and has dimension $\mathcal{D} \approx$ 2. The debate shows the need for
complementary approaches using other data and methods.
 
Recently, Di Nella et al.\ (1996)
analyzed the correlation properties of the galaxy space
distribution using the LEDA all-sky data base containing over
36000 galaxies with known redshifts.  They concluded that the
distribution has a fractal character up to the scales of 300~Mpc
with the fractal dimension $\mathcal{D} 
\approx$ 2.2. In a recent discussion, Guzzo (1997) concluded that
the galaxies are clustered in a fractal way on small and intermediate
scales, and the turnover to homogeneity may occur somewhere
between 200 and 400 Mpc.
Taken at face value, such results predict that in an all-sky sample
like KLUN, one should see a radial
galaxy number density behaviour which deviates from the uniform
distribution.  In terms of the distance modulus $\mu$, uniformity
corresponds to the differential law $\log{N} = 0.6 \mu + \mbox{constant}$ 
for the number $N$ of galaxies in the distance modulus range $\mu \pm 
1/2 \mu$, when $\mu$ has been derived by the ``standard candle'' method.  In a
distribution with fractal dimension $\mathcal{D}$, the expected density law
around any galaxy is $\log{N} = 0.2  \mathcal{D} \mu + \mbox{constant}$, or with 
$\mathcal{D} =$ 2, we have the ``$0.4 \mu$-law''.

There are a few noteworthy aspects in the present study. We concentrate on  
investigating how the radial density behaves around us, while the analyses
mentioned above are usually based on the average behaviour around the 
different galaxies in the sample, which reveals the actual fractal law.
However, there is special importance in seeing directly the behaviour
around our individual vantage point. One thing is that we are naturally eager
to see and check in concrete manner the ``strange'' prediction of fractality,
or the thinning of average density, if our Galaxy occupies an average position 
in the fractal structure. On the other hand, Sylos Labini et al.\ (1996)
have pointed out that in order to  see the fractal behaviour around
one point, one needs an all-sky sample with a sufficiently faint
magnitude limit ($m \ge 14$) in order to avoid so-called finite-volume
effect inside a few Voronoi-lengths. The KLUN sample can thus be used as a
kind of first firm step for deriving the radial density law, which may be
later straightforwardly extended deeper when the sample is gradually
increased.

In order to derive the density law, we use a new method based on  
photometric (TF) distances. It has two special advantages. First,
we can use almost the whole sample for constructing the density law.
Second, unlike the methods based on redshift distances, we do not
require a local velocity field model. In view of the recent suggestion
(Praton et al.\ 1997) that working in the redshift space may enhance or even 
produce structures of the ``Great Wall'' type (and hence influence the 
correlation studies and presumably, the derived value of $\mathcal{D}$), it is
especially important to use photometric distances in complementary studies.

\section{The sample and the method}

\subsection{KLUN}
  
   The KLUN galaxy sample and the quality of the measured
quantities of its members have been recently described by
Theureau  et al.\ (1997b).  Originally intended to be complete
down to the isophotal diameter $1'.6$, it now consists of 5171
galaxies covering the type range Sa -- Sdm, having redshifts and
the measured 21cm-line widths for the Tully-Fisher relation.  4577
galaxies have also the $B_{\rm T}$ magnitude.

   For calculation of distance moduli $\mu$, we use the direct TF relations
established by Theureau et al.\ (1997b) and the same corrections
for type effect, internal extinction, and galactic absorption.
For example, the diameter TF relation gives the distance 
modulus $\mu$ as 
\[ \mu = 5(\log{D_{25}}-\log{d_{25}^{\:0}}+\mbox{const.}), \]
where $\log{D_{25}}=a \logVM + b(T)$. In this formula, 
$d_{25}^{\:0}$ is the angular diameter corrected for galactic
absorption and internal extinction, $V_M$ is the maximum rotational
velocity measured from the 21cm-line width of the galaxy,  
$b(T)$ is the TF zero point for
galaxy type $T$, and $a$ is the slope of the TF relation.

\subsection{The method of TF distance moduli}

   We apply here a method outlined in Sect.\ 3 of Teerikorpi (1993), 
where the TF distance moduli can be used, in principle, up to the 
limits of the sample, in order to derive the form of
the density law.  The method utilizes galaxies in narrow $\logVM$
ranges to construct several differential distributions of $\mu$, 
one for each $\logVM$ range.  These distributions are displaced relative
to each other in $\mu$, because of different $\logVM$ (absolute
magnitude) and the common diameter limit (or more exactly, the
common incompleteness law).  They have also generally different
amplitudes, because different $\logVM$ ranges are populated by
different numbers of galaxies, though their forms are similar if
inside the ranges the galaxies are evenly distributed.

\begin{figure}
\epsfig{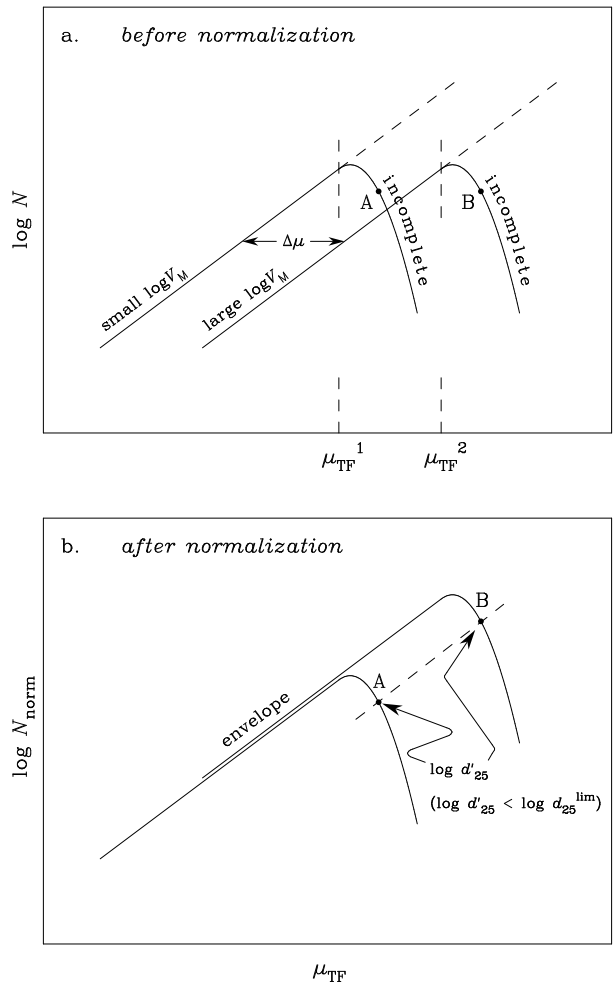}
\caption[]
{Schematic illustration of how the distribution of TF distance moduli 
$\mu_{\rm TF}$ differ for two narrow $\logVM$ ranges separated from each
other by $\Delta \logVM$. Here it is assumed that the true space 
distribution produces a constant slope in the $\log{N}$ vs.\  $\mu_{\rm TF}$
diagram (corresponding to density law $\rho(r) \propto r^{-\alpha}$).
Small-$\logVM$ subsample reveals the correct density law up to 
$\mu_{\rm TF}^{\;1}$ where the incompleteness starting at $\log{d_{25}^{\rm
lim}}$ begins to affect. Large-$\logVM$ subsample shows the same behaviour,
though shifted towards larger $\mu_{\rm TF}$ by $\Delta \mu =
5 a \Delta \logVM$. In the present method such distributions are in 
addition shifted vertically so that the complete parts come together
forming a common envelope which follows the true space distribution law. 
Points A and B correspond to the same angular size $d_{25}'$ in the 
incomplete region. After normalization they also define the correct
density gradient.
}
\end{figure}

\begin{figure}
\epsfig{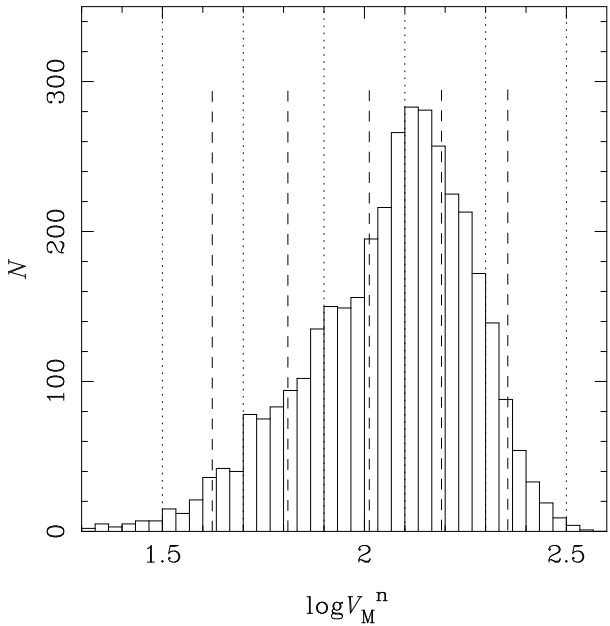}
\caption[]
{$\logVM^{\rm n}$ distribution of the KLUN sample. Bins are separated 
with dotted lines. 
The dashed lines show the average values in each bin.}
\end{figure}

   One expects that above some $\log{d_{25}}$, the sample is complete,
hence below certain limits $\mu_{\rm lim}(\langle \logVM \rangle)$
 each distribution reflects the true average space distribution.
We illustrate this with Fig.\ 1 (see also Fig.\ 1 of Teerikorpi 1993), 
which helps one to understand why division of the sample into 
$\logVM$-ranges allows one {\em to separate the effect of true space
density distribution from the effect of incompleteness in the
distribution of $\mu$.  }
After having identified this $\mu$-range for the subsample with smallest 
$\langle \logVM \rangle$, hence smallest $\mu_{\rm lim}$, one can normalize 
the other subsamples to this one, shifting their $\log{N}$ vs.\ $\mu$ 
distributions vertically, so that up to $\mu_{\rm lim}$ all have the 
same average normalized number counts.  This is allowed because for the 
subsamples with larger $\langle \logVM \rangle$, the complete $\mu$-ranges 
extend beyond the $\mu_{\rm lim}$ corresponding to smaller 
$\langle \logVM \rangle$.

   The common envelope of thus normalized distributions should
follow the space distribution law up to the $\mu_{\rm lim}$ for the
subsample with the largest $\langle \logVM \rangle$.  An interesting 
feature of the method is that one may construct the space distribution still
farther, as is shown by Fig.\ 1b where the points A and B corresponding 
to the same value of $\log d_{25}'$ ($< \log d_{25}^{\rm lim}$) in the
incomplete region are connected by the line having the correct space
density slope.  The method is also
quite simple in essence and allows one quickly to ``see'' the run
of the density law and the factors influencing its derivation. 
Of course, working with real data is more problematic than the
ideal theoretical case.  For instance, one cannot use
indefinitely small $\logVM$ ranges.

\subsection{Special aspects}

   Actually,  one must use the normalized $\logVM$ when making the
described division of the sample:
\[
\logVM^{\rm n} = \logVM - (\log{d_{25}^{\:0}} - \log{d_{25}})/a + (b(T) - b(6))/a
\]
The concept of normalization has been introduced and discussed
in a whole series of studies connected with the so-called method of 
normalized distances (see e.g.\ Theureau et al.\ 1997b). In this
manner, galaxies of different types and with different
inclination and galactic absorption corrections, but with equal
normalized $\logVM^{\rm n}$, have the same effective limiting diameters. 
The $\mu_{\rm TF}$-distributions become thus better defined, and more
closely follow the ideal distribution determined by the space
density and incompleteness curves.  In other words, for a given
$\logVM^{\rm n}$, the $\mu_{\rm TF}$ scale reflects faithfully the 
$\log{d_{25}}$ scale.

\begin{figure}
\epsfig{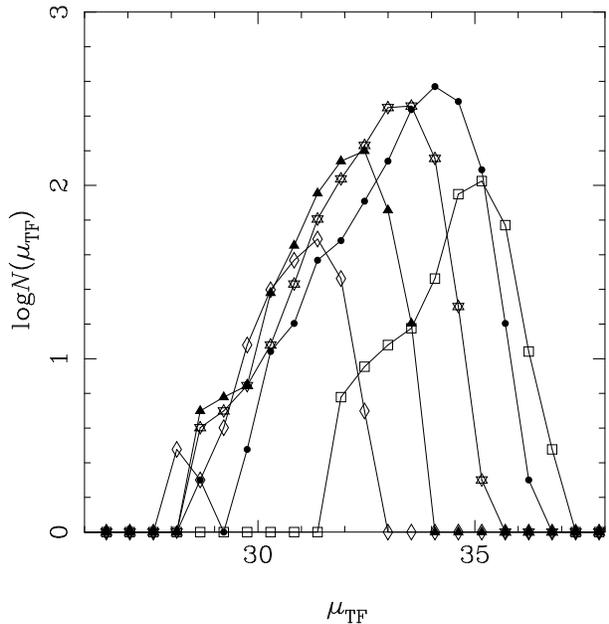}
\caption[]
{Differential $\log{N}$ vs.\ $\mu_{\rm TF}$ (diameter) distributions for
different $\logVM^{\rm n}$ ranges (see the text).}
\end{figure}

Finally, we note that the method is in principle insensitive to the
``patchy incompleteness'' (Guzzo 1997) that may be present in such 
somewhat non-uniform collections of data as LEDA (the host database
of KLUN which, however, was intended to be a homogeneous part of LEDA): 
In different regions of the sky the limiting diameters
(magnitudes) may be different. Let us consider two regions, 1 and 2,
where incompleteness starts at $\log{d_{25}^{\;1}}$ and
$\log{d_{25}^{\;2}}$, respectively, and suppose we cannot identify
these regions from the data. The summed relation, $\log{(N_1+N_2)}$
vs.\ $\mu$, for a constant $\logVM$ still follows the correct
slope at small distances ($N_1 + N_2 =N_1 + c \cdot N_1 = (1+c) N_1$,
where the constant $c$ depends only on the ratios of limiting
angular sizes and region areas) while now the incomplete part of the curve
is deformed. However, for different $\logVM$ the form of the curve is
similar and the situation is analogous to what is depicted by Fig.\ 1.

\section{Search for the density law}
   
\subsection{Diameter TF distance moduli}

   We show the procedure in a transparent manner which allows one
to see the steps taken and to recognize the impact of possible
systematic errors.

   First, we have divided the sample into five $\logVM$ ranges
according to the normalized $\logVM^{\rm n}$: 1.5 -- 1.7 -- 1.9 -- 2.1 -- 
2.3 -- 2.5.  These intervals cover practically all the sample and the
middle one is centered on the maximum of the $\logVM$ distribution
(Fig.\ 2). Inside the extreme ranges the distributions are not symmetric
around the mean, due to the systematic effect described in Sect.\ 3.3.

\begin{figure*}
\epsfig{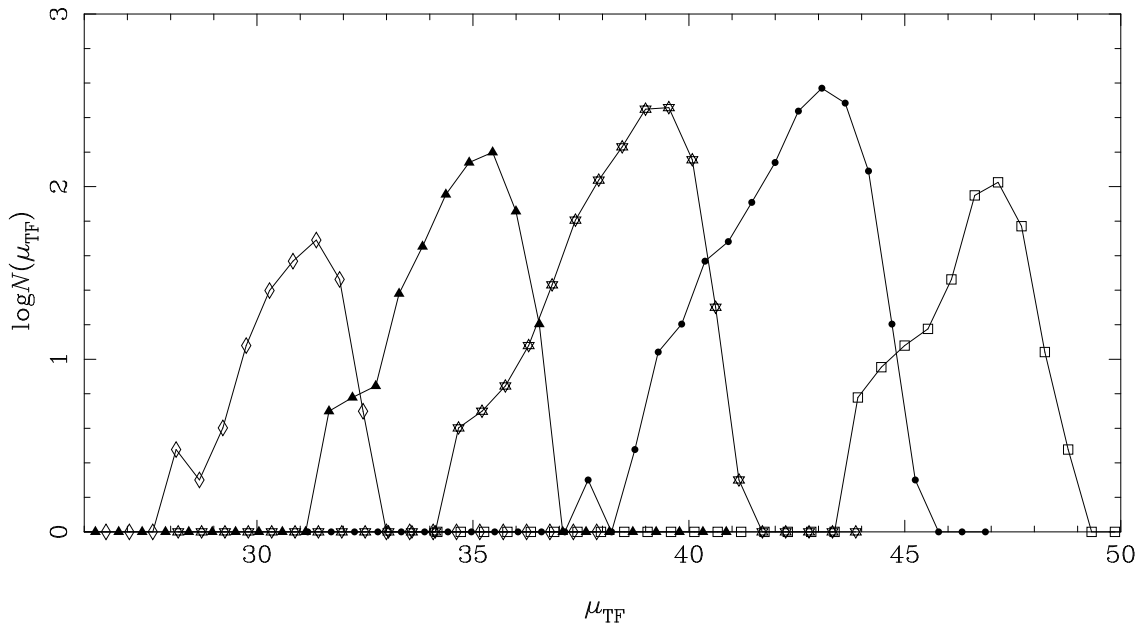}
\caption[]
{As in Fig.\ 3, but now the distributions are arbitrarily
shifted on the $\mu_{\rm TF}$-axis.}
\end{figure*}

   The distributions of $\mu_{\rm TF}$ in the $\logVM^{\rm n}$ 
intervals are shown in Fig.\ 3.  The bins on $\mu_{\rm TF}$ should 
have the width $5 \cdot 1.082 \cdot 0.2 / k$, ($k=$1,2,3,...) in order
to  collect the same  $\log{d_{25}}$ ranges on the individual 
$\mu_{\rm TF}$ distributions as expected from the $\logVM$ intervals (0.2) 
and the slope of the diameter TF relation (1.082).  Selecting $k=2$,
we get suitable bin size $\Delta \mu = 0.541$. For clarity, Fig.\ 4 
shows the distributions artificially shifted on the $\mu_{\rm TF}$-axis, 
revealing their similar forms.

   The next step is the normalization of the numbers $\log{N}$.  For
this we use the following procedure:
\begin{enumerate}
\item Select the complete part of the $\log{d_{25}}$ distribution.  The
putative limit is $\log{d_{25}} = 1.2$, as expected from previous
studies.
\item Find the corresponding part of the first subsample (with
smallest $\logVM^{\rm n}$) and count the total number of galaxies below 
and at $\mu_{\rm lim}(1) = 30.5$: $N(1,\mu_{\rm lim}(1))$.
\item For the second subsample calculate $N(2, \mu_{\rm lim}(1))$.
\item Shift curve 2 by $\Delta \log{N} = \log{N(1, \mu_{\rm lim}(1))}
 - \log{N(2,\mu_{\rm lim}(1))}$.
\item Normalize the remaining curves using the same procedure with
increasing $\mu_{\rm lim}$ by the shift between the subsequent
curves.  In this manner we utilize progressively deeper complete
parts.  Then the cumulative correction for the j:th curve is
\begin{eqnarray*}
\Delta_j \log{N} = \sum_{i=2}^j & \left[  \log{N(i-1, \mu_{\rm lim}(i-1))} 
\right. \\
 & \left. - \log{N(i, \mu_{\rm lim}(i-1))} \right] ,
\end{eqnarray*}
where $\mu_{\rm lim}(i) = \mu_{\rm lim}(i-1) + 0.541$. By restricting 
the steps in this manner, we keep progressively farther from the putative
limit $log{d_{25}} = 1.2$, while having an increasing number of galaxies
available for normalization.
\end{enumerate}

   Figure 5 shows the resulting normalized composite distribution. 
One may note two particular things:  There is first a rather
scattered run of points up to about $\mu_{\rm TF} = 31$ which can be roughly
approximated by a $0.6 \mu_{\rm TF}$ line.  After that the common envelope
assumes a shallower slope of about 0.46.

   Using now the points above $\log{d_{25}} = 1.2$  for each curve and
requiring that each point contains more than 12 galaxies, by
which the noisy $\mu_{\rm TF}$ ranges are removed, we draw a straight line
through them.  For this line, there are points up to $\mu_{\rm TF} = 34$. 
We use this line, derived by simple least-squares technique, as a
reference for further study of the density law using information
from the incomplete parts of the $\mu$ - distributions.   This
``envelope line'' is shown in Fig.\ 6 together with points from the
incomplete parts where we again use only those which contain more
than 12 galaxies  The average $\langle \log{d_{25}} \rangle$
is written besides each symbol.  One can easily discern the sequences 
corresponding to the same positions in the $\log{d_{25}}$ distributions
(c.f.\ points A and B in Fig.\ 1).  
These averages do not vary much along the same sequence.  Comparing 
with the line of slope 0.46 inserted, one sees that the sequences follow
rather well this slope, similarly as was found above for the points in 
the complete part (especially above $\mu_{\rm TF} \approx 31$).

\begin{figure}
\epsfig{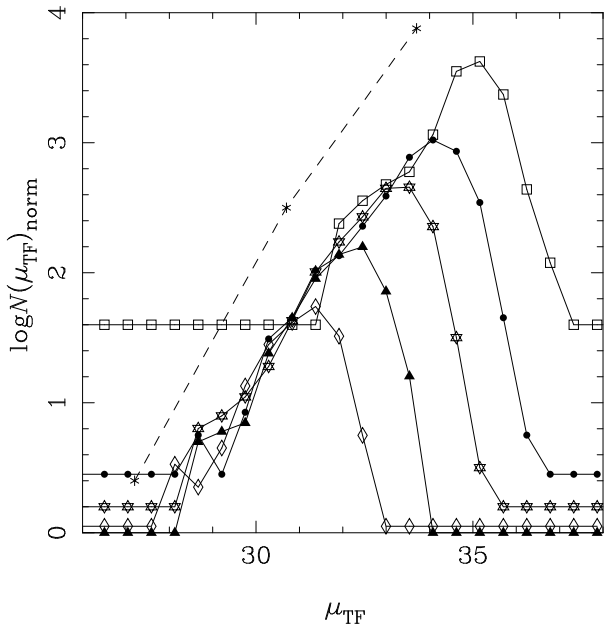}
\caption[]
{Normalized composite distribution, constructed from the
distributions of Fig.\ 3 by the steps described in the text.  Note
the appearance of a good envelope line.  The inserted dotted
lines have slopes 0.6 and 0.46.  $\log{N} = 0$ levels are conveniently
identified as horizontal parts.}
\end{figure}

\begin{figure}
\epsfig{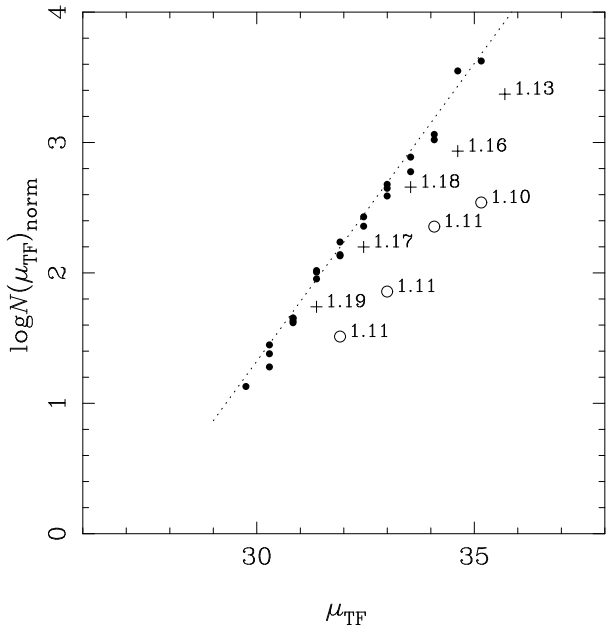}
\caption[]
{The envelope line from Fig.\ 5 (shown as dotted line),
together with points from the incomplete parts.  The dots
represent the envelope points and other symbols refer to
incomplete parts.  The numbers besides the latter give $\langle
\log{d_{25}} \rangle$ and allow one to easily recognize 
the sequences following a slope close to 0.46 (envelope line).}
\end{figure}

\begin{figure}
\vspace*{.4cm}
\epsfig{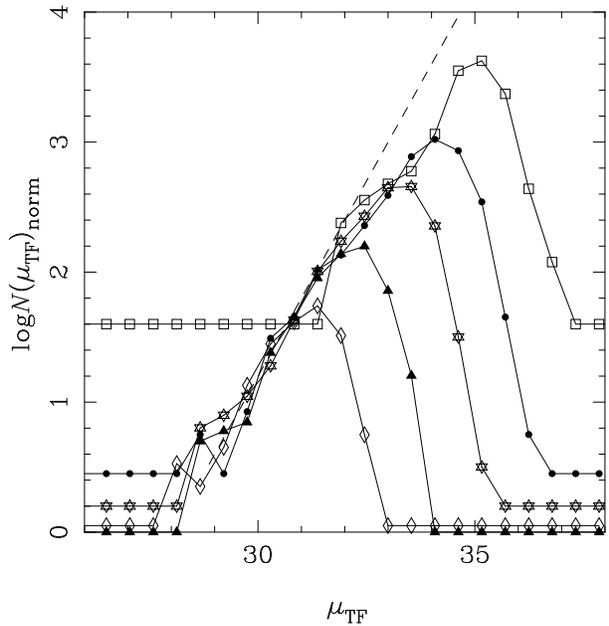}
\caption[]
{The line of slope 0.6 forced to go through the first
normalization point.  In this case the incompleteness is seen to
increase together with $\logVM$.}
\end{figure}

We tested the completeness limit by assuming 
limits  at  $\log{d_{25}} =$
1.3 and 1.4 instead of the above  $\log{d_{25}}=1.2$. With these more
conservative values we can be more certain of the completeness of
the sample, however at the price of reduced number of data points.
We got slopes 0.46 ($\pm 0.01$) and 0.43 ($\pm 0.02$) for limits 1.3
and 1.4 respectively, but the ``envelope lines'' in both cases
were not as well defined as before. 
For now, we are more satisfied with the completeness limit
at $\log{d_{25}}=1.2$ giving us the slope 0.46 (see also Sect.\
3.4 for another test, using simulated galaxy samples).

   Naturally, all this is interestingly close to 0.44 predicted on
the basis of fractal dimension $\mathcal{D} \approx$ 2.2, as obtained e.g\ in
the Di Nella et al.\  (1996) correlation analysis of LEDA.  
However, we are well aware that systematic effects may be
involved, when we are working rather close to the completeness
limit of the KLUN-sample.  We have attempted to find any
systematic error that could explain the shallower than 0.6 slope
and have also looked what happens when one forces a line of slope
0.6 to start at $\mu_{\rm TF} = 31$.  In Fig.\ 7 this has been done, and the
line which rather well describes the (scattered) run of data
below $\mu_{\rm TF} = 31$, systematically deviates from the envelopes of the
normalized distributions at larger $\mu_{\rm TF}$.  This behaviour implies
that if the deviation from the $0.6 \mu$-law is due to incompleteness
problem, the incompleteness in apparent size starts for different $\logVM$ 
ranges at different $\log{d_{25}}$.  This is something that we
cannot understand, because the needed effect is quite large.  For
the largest $\logVM$ it means that incompleteness starts around $\log{d_{25}}
 = 1.7$, while for the smallest $\logVM$ such a limit would be 
around 1.2 (in terms of magnitudes this corresponds to a
difference of about 2.5 mag).  Still another way to state the
problem is that if we try to shift the maxima of the
distributions to follow the $0.6 \mu$-slope, the curves are everywhere
separated, there is no normalization and no common envelope.  A clear 
argument against such a large effect comes also from the method of 
normalized distances used in Theureau et al.\ (1997b): in the 
``unbiased plateau'' produced as a part  of the method, one 
should readily recognize such differences in the selection
functions of galaxies with small and large $\logVM$.  Also, from the
manner of how KLUN was created as a diameter limited sample,
independent of any considerations of $\logVM$, there is no reason to
expect so significant dependence of the selection on $\logVM$.  For
instance, a recent analysis of the H {\sc i} line profile detection
rates at Nan\c{c}ay radio telescope by Theureau et al.\ (1997c) does
not give any indication that large $\logVM$ galaxies are significantly
underrepresented in KLUN.

   Finally, if erroneous, the present slope 0.46 is just by an 
accident very close to ones obtained by several quite different 
correlation analyses (e.g.\ Sylos Labini et al.\ 1998a).

\subsection{Magnitude TF distance moduli}

   The $B$-magnitude TF relation has smaller scatter than the
diameter relation, which makes it tempting to use also it in this
study even though the numbers are then smaller and the selection
properties of the resulting sample are complicated by the fact
that KLUN has been originally selected on the basis of apparent
size.  However, if the distribution of the galaxies of different
$\logVM$ ranges is the same in the $\log{d_{25}}$ -- $B_T$ plane, one can use
the magnitude relation in a similar manner.

   We do not go through the steps in such detail as for the
diameter distance moduli.  Figure 8 shows directly the normalized
composite diagram.  Because of the underlying diameter limit, the
incompleteness begins farther from the maxima than in the case of
diameters, and the complete part of the envelope is now shorter
in $\mu_{\rm TF}$.  The envelope line constructed from points below the
putative magnitude limit $B = 13.25$, is also shown. It has now the
slope 0.40.  Because the numbers of galaxies at each point
of this diagram are smaller than in Fig.\ 5, the error of the slope
is slightly larger than when using diameters ($\sigma=0.012$, compared
to $\sigma=0.010$ with diameters). Inspection of the points in the
incomplete part shows that the sequences, analogous to what was
discussed for the diameter moduli, have similar slopes.  However,
now the achievable $\mu$-ranges and numbers are smaller, and we do
not show them separately. 

\begin{figure}
\epsfig{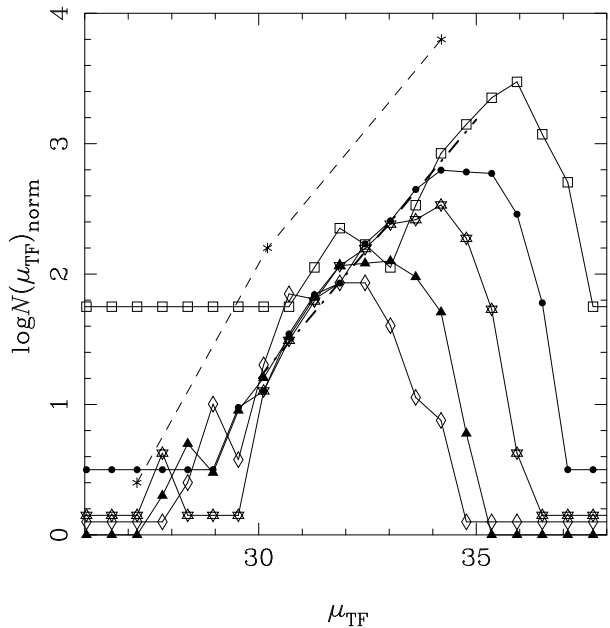}
\caption[]
{Normalized composite distribution for $B$-magnitude TF
distance moduli in different $\logVM^{\rm n}$ ranges.  
The slopes 0.6 and 0.4 are shown as well as the envelope line.}
\end{figure}

\begin{figure}
\epsfig{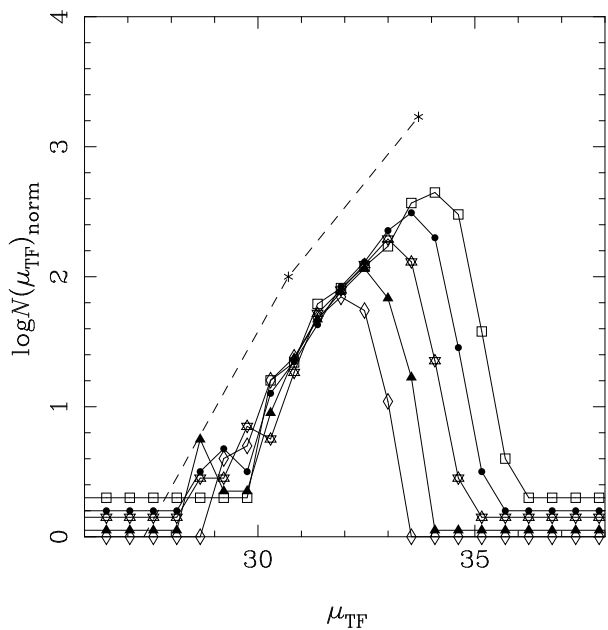}
\caption[]
{Normalized composite distribution for diameter TF distance
moduli, constructed as Fig.\ 5, but using intervals of 0.1 in
$\logVM^{\rm n}$ in the range 1.7 -- 2.3.  Note the appearance of a good
envelope line.  The inserted dotted lines have slopes 0.6 and 0.41.}
\end{figure}

\subsection{Systematic error caused by finite $\logVM$ ranges}

   One can see a systematic effect in this method where we are
forced to use finite $\logVM$ ranges instead of ideal
infinitesimals.  It comes from the fact that galaxies are not
quite similarly distributed inside the different $\logVM$ ranges.  
The distribution of $\logVM$ has a maximum.  Because of the form of
the $\logVM$ distribution, inside the small - $\logVM$ interval one
expects an increasing number density of galaxies towards the edge
with larger $\logVM$, while for the large - $\logVM$ range this trend
is reversed (see Fig.\ 2). In Fig.\ 6 one sees that the
averages $\langle \log{d_{25}} \rangle$ within different sequences 
do not vary very much which suggests that the effect is actually not very
important.

   In order to check whether decreasing the interval size
influences the result obtained above, we made an experiment
whereby $\logVM^{\rm n}$ interval was reduced to 0.1, in the range 1.7 --
2.3, where the numbers of galaxies remain large enough.  Now the slope
of the envelope line is 0.41, the furthest point of the line being
slightly above 100~Mpc (Fig.\ 9). Again, the diminished number of
galaxies make error of the slope larger, $\sigma=0.015$, while 
$\sigma=0.010$ in Fig.\ 5.

\subsection{Numerical experiment using simulated galaxy sample}

In order to check further the reliability of the used method, we have
made numerical experiments with simulated galaxy samples. 
In making these tests, we have kept in mind the following points:
\begin{enumerate}
\item Because individual distance moduli have considerable scatter
and because we are interested in the all-sky averaged behaviour
of radial density, the present method and the available data
allow one to derive a quite smoothed-out view of the space
density around the Galaxy.
\item The all-sky average, as derived by the present method, does
not make a difference between a random distribution of galaxies
with a radial density variation and a fractal distribution with
the corresponding fractal dimension.  Hence, for the purpose of
testing systematic errors in the method, it is sufficient to
consider simulated distributions, where randomly scattered
galaxies have a smooth radial density variation.
\item The all-sky averaged radial distribution inside a fractal
structure is statistically the same around all galaxies, which
gives special motive for applying the present method.  In
non-fractal structures, such as supercluster-void network (e.g.\
J.\ Einasto et al.\ 1997), the radial distribution depends on the
position of the observer (though in many cases the presence of the plane 
gives an apparent $\mathcal{D} \approx$ 2, also reflected in correlation
function analysis where actually an average of all the observers
is taken, see the models by Einasto 1992).  It is intended to
extend the present method to study large individual structures in
specified regions of the sky detected previously with
redshift-distances (such as the Great Wall).  Such applications
will need specially tailored simulations which show how the
method draws the density distribution curve, say, through a
narrow plane of galaxies.
\end{enumerate}

\begin{figure}
\epsfig{file=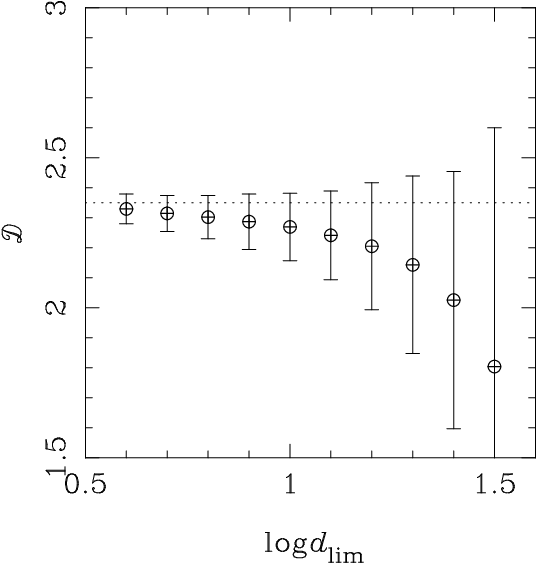,angle=270,width=8cm}
\caption[]
{Test of the apparent diameter completeness limit  
$\log{d_{\rm lim}}$ using simulated galaxy samples. Dotted line shows
the input value of density gradient $\alpha=0.47$, which corresponds
to fractal dimension $\mathcal{D} =$ 2.35. For more complete samples
(smaller $\log{d_{\rm lim}}$) the observed  $\mathcal{D}$
approaches the input value. Error bars are mean deviations ($2 \sigma$) for
1000 simulations. $\log{d_{\rm lim}}=1.2$ 
is the completeness limit assumed for KLUN. }
\end{figure}

In the experiments we started with large number of galaxies (say $10^5$),
for which we alloted radial distances using an input value of the
radial density gradient
$\alpha=\alpha_{\rm in}$ in the density law $\log{N}=\alpha \mu +$ const. 
For each galaxy we chose a random absolute diameter from  a gaussian
distribution with $\log{D_{\rm abs,0}}=1.3$ and $\sigma=0.18$
($D_{\rm abs}$ in kiloparsecs). Even though the $\log{D_{\rm abs}}$
distribution in reality hardly is gaussian, it is not too far
from the truth for the KLUN sample. All the galaxies having apparent diameter 
larger than limit 
$\log{d_{\rm lim}}$ were included in the ``observed subsample''.
To make this observed sample more realistic we also allowed in some galaxies
below the diameter limit, percentage of included galaxies being progressively
smaller further below the $d_{\rm lim}$. Each of these observed galaxies
were then  given a rotational parameter $\logVM$ by inverse TF relation 
($\logVM = a' \log{D_{\rm abs}} + b'$) with $a'=0.5$, $b'=1.13$
and a gaussian dispersion $\sigma=0.15$. These values resemble 
results of the recent KLUN study (Theureau et al.\ 1997a). Now the
$\log{N}$ -- $\mu_{\rm TF}$ graph (distance moduli from {\em direct}
TF relation) for the observed subsample could be investigated 
as above was done for the real KLUN sample.

First we selected $\log{d_{\rm lim}}=1.2$ and varied the density gradient
$\alpha_{\rm in}$. For each $\alpha_{\rm in}$ we chose the initial number of galaxies
so that the number of observed galaxies was the same as in KLUN sample.
Then we calculated the slope of the ``envelope'' line in the $\log{N}$
-- $\mu_{\rm TF}$ graph getting 
the observed density gradient $\alpha_{\rm obs}$
For each $\alpha_{\rm in}$ we repeated the simulation 1000 times 
to get $\langle \alpha_{\rm obs} \rangle$ with error bars.
For every $\alpha_{\rm in}$ the resulting $\langle \alpha_{\rm obs} \rangle$
was smaller than $\alpha_{\rm in}$. However, this tendency 
was not large enough to explain the deviation of our observed
slope (in this section we use as a reference point $\alpha=$ 0.44, 
which is a weighed average of the slopes
obtained with diameters and magnitudes) from homogeneous galaxy 
distribution ($\alpha=$ 0.6). The simulations showed that for
$\alpha_{\rm obs}=$ 0.44, $\alpha_{\rm in}=$ 0.47 $\pm 0.04$ 
($2 \sigma$ errors). In terms of fractal dimension ($\alpha =
0.2 \mathcal{D}$) we can say that the observed value
$\mathcal{D} =$ 2.2 can be affected by our methods so that
the true value is $\mathcal{D} \approx$ 2.35 $\pm 0.20$.

We then tested the effect of the completeness limit by varying 
$\log{d_{\rm lim}}$, while keeping the other parameters fixed. 
Figure 10 shows how $\mathcal{D}_{\rm obs} 
\rightarrow \mathcal{D}_{\rm in}$ when the sample gets more complete.
Also, with smaller completeness limits the number of observed 
points increase, and the errors get smaller.
This emphasizes the importance of expanding the database to make
it deeper and more complete.

\section{Discussion and concluding remarks}

   As it appears from the preceding analysis, we have arrived at
the conclusion that KLUN galaxies seem to indicate a ``thinning''
of the galaxy universe from $\mu_{\rm TF} \approx 31$ up to $\approx 36$.  
In order to see the actual range, the distance moduli must be  corrected 
for the Malmquist bias of the first kind which in the case of $\mathcal{D} =$ 2
becomes $\mathcal{D}/\mbox{3} \times$ 1.382 $\sigma_M^2 = 0.92 \sigma_M^2$ for the magnitudes and  
$\mathcal{D}/\mbox{3} \times$~$5 \times 5 \times 1.382 \sigma_{\log{D}}^2 = 
23.05 \sigma_{\log{D}}^2$ for the diameters (see e.g.\ Teerikorpi 1997).
Using $\sigma = 0.5$ mag and $= 0.15$ for magnitudes and (log)diameters, 
respectively, these corrections make the maximal distance probed by both 
the diameter and magnitude relations about 200~Mpc, corresponding to the 
first point beyond the maximum of the highest $\logVM$ distribution.

This result, using photometric distances independently of redshifts,
and an all-sky sample,  offers a complementary piece of evidence for the debate
on the dimension and maximum scale of fractality. The value $\mathcal{D}
\approx$ 2 which has appeared in several correlation studies 
(Klypin et al.\ 1989, Einasto 1991, 1992, Guzzo et al.\ 1991, Calzetti et al.\
1992, Di Nella et al.\ 1996, Guzzo 1997, Sylos Labini et al.\ 1998a), is here seen as a density law
up to 200 Mpc, the limit of the KLUN sample. Taken alone, the present
study does not prove any large scale fractality, it only suggests 
the existence of such an average density law around us 
(which in a restricted distance range could be caused by filamentary 
structure and voids or a large flattened system; Einasto 1992, Einasto et al.\
1994, Paturel et al.\ 1994). On the other hand, within the framework of a large
scale fractal distribution, our position on  a peak surrounded by
a decreasing density is as expected (Mandelbrot 1982).
In this sense, when considered together with large scale correlation studies 
leading to $\mathcal{D} \approx$ 2, 
the present result is consistent with the view that the Milky Way
occupies a typical position in the fractal structure.
This agrees with recent studies by Karachentsev (1996) and Governato et al.\
(1997) who conclude that the Local Group is a typical representation
of other small groups, both in structure and environmental dynamics.

The result of our work is also consistent with recent results on the
supercluster--void network (M.\ Einasto et al.\ 1994, 1997) and arguments
based on the correlation analysis of rich clusters of galaxies (Einasto \&
Gramann 1993, J.\ Einasto et al.\ 1997). These studies claim that the 
observed transition scale to homogeneity is about 200--300 Mpc.

In view of local structures, fractal or not, we would not
expect any perfect $0.6 \mu$-law. Nevertheless, we have tried to
understand if such a strong deviation from a homogeneous distribution
could be due to some systematic error. However, numerical tests have
shown that the method works, though it tends to somewhat underestimate
the density gradient.

Naturally, the present study must be seen as a first step
which uses distance moduli in the described manner and the method must
be developed along with the increasing size of KLUN. A promising 
prospect is offered by the ongoing DENIS project (collecting 
200000 galaxies by 1999) and the FORT project improving the
efficiency of the Nan\c{c}ay radio telescope (see Theureau 1997).
A deeper sample will allow us to better recognize systematic errors and the
real space density trend.

\acknowledgements{
We are grateful to Prof.\ Jaan Einasto for his very useful critisism.
M.H. thanks for support by Finnish Culture Foundation, M.\ Ehrnrooth's
Endowment and A.\ Kordelin's Cultural and Educational Foundation.
Yu.B.\ thanks for support by the Russian program
``Integration'' project N.\ 578. 
This work has been supported by Academy of Finland (project ``Cosmology
in the Local Universe''). 
We have used data from the Lyon-Meudon Extragalactic
Database (LEDA) compiled by the LEDA team at the CRAL -- Observatoire
de Lyon (France).}

\end{document}